# Micromagnetism in (001) magnetite by spin-polarized low-energy electron microscopy

Juan de la Figuera[1], Lucía Vergara[1], Alpha T. N'Diaye[2], Adrian Quesada[3], and Andreas K. Schmid[2]

1 Instituto de Química-Física "Rocasolano", CSIC, Madrid 28006, Spain
2 Lawrence Berkeley National Laboratory, Berkeley, California 94720, USA
3 Instituto de Cerámica y Vidrio, CSIC, Calle Kelsen 5, 28049, Madrid, Spain

**Abstract**

Spin-polarized low-energy electron microscopy was used to image a magnetite crystal with (100) surface orientation. Sets of spin-dependent images of magnetic domain patterns observed in this surface were used to map the direction of the magnetization vector with high spatial and angular resolution. We find that domains are magnetized along the surface [110] directions, and domain wall structures include 90° and 180° walls. A type of unusually curved domain walls are interpreted as Néel-capped surface terminations of 180° Bloch walls.

**Highlights**

- The (100) surface of magnetite is imaged by spin polarized low energy electron microscopy.
- The magnetic domain microstructure is resolved.
- Magnetic easy axes in this surface are found to be along <110> directions.
- Magnetic domain wall structures include wide Néel-caps.

## 1. Introduction

Magnetite is the most strongly magnetized material found in nature. Magnetized "lodestone" (i.e., magnetite [1]) has been known to humankind at least since 1 BCE in ancient Greece. As the oldest magnetic material, it has played an important role throughout the history of science and technology, and the discipline of paleomagnetism relies to a large extent on its properties. More recently, band theory predicted that magnetite is a half-metal [2,3], i.e. its conduction electrons are expected to be 100% spin-polarized. Interest in magnetite as a candidate material for spintronic applications [4] is driven by the predicted half-metal character, high conductivity, chemical stability and high-Curie temperature. Promising recent results include the use of magnetite as a spin-injector source [5] as well as observations suggesting that the magnetic properties of magnetite are robust even in the nanometer thickness limit [6].

At room temperature (RT) magnetite has a cubic inverse spinel structure which can be visualized as a face centered cubic arrangement of the oxygen anions [7]. Iron atoms are located at sites within tetrahedra ($T_d$ sites) or within octahedra ($O_h$ sites) of oxygen atoms, respectively. While $T_d$ sites are populated with $Fe^{3+}$ cations, the $O_h$ sites have both $Fe^{2+}$ and $Fe^{3+}$ cations. The ferrimagnetic order of magnetite arises from a ferromagnetic coupling between iron in tetrahedral sites and ferromagnetic coupling between iron in octahedral sites, while the crossed $T_d$ and $O_h$ site interaction is antiferromagnetic. The Curie temperature is ~850 K and the easy magnetization directions are the <111> lattice directions [7].

The (001) surface of bulk magnetite has been the subject of many experimental studies aimed at determining its structure, using tools including low- energy electron diffraction [8, 9], X-ray photoelectron diffraction [10], low-energy ion scattering [11, 12], scanning tunneling microscopy [10, 13, 14, 15] and spectroscopy [16]. The most common surface reconstruction, $\sqrt{2}\times\sqrt{2}R45°$, has been alternatively interpreted as octahedral-iron terminated with a Jahn-Teller distortion [17] or as tetrahedral-iron terminated. Recently, it was shown that both terminations can be obtained by modifying slightly the preparation procedure [18]. After experimental work to test the predicted half-metal character had led to inconsistent results [19, 20, 21], the √2×√2R45° reconstruction was suggested as possibly playing a role in suppressing the expected 100% spin polarization at the Fermi level. It was shown that lifting the reconstruction restores the half-metallicity [18, 22].

Much of the experimental work was performed on thin films grown on appropriate substrates. Samples included (100) oriented films grown by reactive molecular beam epitaxy, magnetron sputtering or pulsed laser deposition on MgO [23, 24, 23, 25], $SrTiO_3$[25, 26] and $MgAl_2O_3$ [23, 27, 28], and magnetic domain patterns were observed by magnetic force microscopy [28] and by photoemission microscopy [29]. However, in thin film samples residual strains and antiphase boundaries or other defects may affect magnetic properties [23, 28] and it is interesting to compare with measurements on bulk crystals. Magnetic domain structures on some surfaces of bulk magnetite crystals have been studied using the Bitter technique [30], Kerr-effect Microscopy [31, 32], magnetic force microscopy [33, 34], and holography in transmission electron microscopy [35].

The micromagnetic structure of the (001) surface of bulk magnetite crystals is less well known. The (001) surface is not aligned with any bulk easy-axis directions, and intricate patterns including closure domain features and quasidomains are thus expected [36]. Goal of this work is to reveal such features in detail by mapping the direction of the magnetization vector on the (001) surface of bulk magnetite.

2. Methods

Spin-polarized low-energy electron microscope (SPLEEM) [37] is a tool that permits measurement of the components of the magnetization vector along real-space coordinates x, y, and z. In this instrument a beam of polarized electrons is reflected from a crystalline sample. The spin-polarized electron gun contains optics which permits electron energy as well as the quantization axis of the electron beam spin-polarization to be freely adjusted [38]. The electron reflectivity of the sample surface varies as a function of energy and, in case of magnetic surfaces, an additional (smaller) spin dependent contribution modulates the reflectivity. To measure this smaller magnetic modulation, the spin-dependent reflectivity difference is usually expressed as an asymmetry signal $A = (I_+-I_-)/(I_++I_-)$, where $I_+$ and $I_-$ are reflectivities for electron beams with spins aligned and anti-aligned with a chosen quantization axis. Two full-field reflectivity images are acquired with opposite spin direction of the electron beam and pixel-by-pixel "asymmetry" images are computed. Since the asymmetry signal $A$ is proportional to the scalar product of the sample magnetization vector and the spin quantization axis of the electron beam [39], brightness (darkness) in an asymmetry image indicates the magnitude of the local surface magnetization component parallel (antiparallel) to the spin-polarization direction of the electron beam. One way to achieve more quantitative results is to repeat this measurement three times with the electron beam spin-axis aligned along x, y and z spatial directions. Quantitative comparison of the asymmetry signal (gray-values) in these triplets of images then allows us to map the real space orientation of the magnetization vector in the sample surface with high angular and spatial resolution [37].

The SPLEEM vacuum chamber is equipped with a suite of standard surface-science tools, which were used to prepare a (001)-oriented single-crystal of natural origin [40]. After introduction in the SPLEEM system the sample surface was ablated by several cycles of 10 minutes sputtering with Ar ions at 1 keV followed by annealing to 600 °C in $10^{-6}$ Torr of $O_2$ for tens of minutes, until Auger spectroscopy showed only iron and oxygen, and the low energy electron diffraction pattern showed a sharp √2×√2R45° pattern that has been attributed to the clean surface [9, 17].

## 3. Results

To determine optimal settings of the electron beam illumination for magnetic domain imaging, we start by measuring the energy dependence of the electron reflectivity and the energy dependence of the asymmetry signal; the spectra obtained from the freshly prepared √2×√2R45° magnetite surface are plotted in Fig. 1(a) and Fig. 1(b), respectively. Signal-to-noise in SPLEEM images scales with the product of the square of the asymmetry and total reflectivity; for this work we found that the combination of good asymmetry and reasonably large reflectivity in the energy region around 4.7 eV yields good image contrast in relatively short image integration times.

In a second step, easy-axes of magnetization were determined by measuring the relative magnitude of the asymmetry signal as a function of the electron beam spin direction. When the spin-polarization is normal to the surface plane, as in the SPLEEM image reproduced in Figure 2 (a), we find that magnetic contrast nearly vanishes (we measure less than 6% of the magnitude of the in-plane asymmetry signal). When the beam spin polarization is adjusted to be in-plane, magnetic domain patterns are observed throughout the surface: the images reproduced in Figure 2 (b) and (c) were recorded in the same sample position as (a), with beam spin polarization oriented along the directions indicated by arrows. In such a triplet of SPLEEM images, recorded with orthogonal beam polarization directions, gray-values of image pixels correspond to the relative magnitude of the magnetization vector in each pixel along a given direction. The data was converted into a color image, shown in panel (d), where the direction of magnetization is represented in color according to the color-wheel shown to the left of the image. As can be seen from this color-image, most parts of the surface are magnetized approximately along the image diagonal directions, which correspond to the in-plane [110] and [1-10] axes of this crystal. The observation of preferred magnetization axes can be confirmed in a statistically more robust way by plotting the magnetization vector measured of each pixel in form of a scatter-plot as shown in panel (e). The plot clearly resolves the two preferred magnetization axes.

A collage of several images shown in Figure 3 surveys typical magnetic domain patterns found on this magnetite (001) surface. Most domains walls are curved, both for 90° and 180° types. There are several detailed observations that stand out, as highlighted in Figure

4. We observe regions where 90° domain walls form periodic arrays, as shown in Fig. 4(d). The 180° domain wall have substantial width of the order of 0.8 micrometers, and the often form a zig-zag pattern [Figure 4(b)]. The orientation of the magnetization vector within these domain walls is in-plane, points at 90° with respect to the bordering domains, and is nearly constant across the width of the wall, as shown in Fig. 4(c). These features indicate that these 180° domain wall are likely Néel-capped sub-surface Bloch walls. This interpretation is consistent also with simulations of Bloch walls close to a surface in magnetite, which have found Néel caps with a thickness of around 70 nm [42]. An example of a Bloch line, i.e. vortex-line boundaries between domain wall segments of opposite chirality, is seen in Figure 4(d).

While curved domain walls have been linked to stress and crystalline defects in experimental observations of micrometer-scale grains[43, 44], stress is less likely to be an important factor in our surface of a 5 mm-wide annealed single crystal. Instead, following [41] we suggest that domain wall curvature in this case arises from the accommodation of the bulk domains aligned along bulk easy-axis directions in combination with the gradual rotation needed to turn the easy-axis directions into surface [110] directions. Features similar to our observations (wavy domain walls, smooth changes in the magnetization direction) have been predicted from micro-magnetic simulations [41] of up to 4 μm wide magnetite cubes. These simulations also showed that, at the same time, the domain walls are straight in the bulk of the grain. We note that neither the micro-magnetic simulations nor our experimental observations have presented any trace of quasi-domains formed by arrangement of small domains that on a local scale have a sizable out-of- plane magnetization which averages out on a larger scale [36]. Within the resolution of the SPLEEM, we find no fine structure with a significant out-of-plane magnetization. The observation of the Néel caps is consistent with simulations of Bloch walls close to a surface in magnetite, which have shown that the thickness of the Néel caps is around 70 nm [42]. We note that while Bloch walls are chiral (a Bloch wall possesses a sense of rotation) inversion symmetry of the crystal structure suggests that overall, the magnetism in magnetite will not show chiral asymmetry. The observation of Bloch lines, i.e. vortex-lines separating sections of opposite chirality within a domain wall (such as the Bloch line seen in Figure 4d), suggests chiral symmetry of the magnetism in magnetite.

The universal in-plane magnetization we find in this surface merits consideration. In bulk magnetite, the easy magnetization directions are the (111)-type lattice directions. These directions are approximately 55° from the normal direction of the (001) surface and both the out-of-plane as well as the in-plane components of magnetization vectors pointed in these directions would be expected to be large. Consequently, one might have expected to observe a strong asymmetry signal both for electron beams polarized parallel to the surface and for beams polarized in the surface normal direction. The almost vanishing out-of-plane signal we observe shows that magnetization in the surface of our crystal is not aligned with the (111)-type easy magnetization directions of bulk magnetite; instead, the easy magnetization axes in the magnetite (001) surface are rotated very close towards the in-plane axes [110] and [1-10]. The fact that magnetization in the surface of our crystal is different from the bulk easy magnetization directions can be understood to be the result of dipolar forces (shape anisotropy) that tend to bend the easy-axis directions towards the surface, aligning the magnetization vector towards the projections of the volume (111) directions onto the (001) surface, which are the surface [110] directions. Our experimental results are consistent with predictions from micro-magnetic simulations in cubic magnetite grains with (001) sides [41], where the magnetization vector was found to be close to [110] in the near-surface regions of up to 0.25 μm thickness. In-plane [110]-type easy magnetization axes have also been found experimentally in thin magnetite films [see for example, (001) magnetite films on MgO [29], or films on $SrTiO_3$ and $MgAl_2O_3$ [23]]. Also in ref [29], fourfold in-plane magnetic anisotropy with easy-axis along the [110] surface directions was detected. As total anisotropy in (001) oriented magnetite samples results from a competition between magnetocrystalline forces (favoring (111)-type orientation of the magnetization) and dipolar forces (favoring in-plane magnetization near the sample surface) some degree of magnetization canting can be expected. In fact, calculations for a (111) surface [30] have suggested that the surface easy-axis would be canted by 2° from the surface plane. A small canting of the magnetization is also compatible with our out-of plane SPLEEM images (not shown) where the asymmetry signal is very weak but does not vanish completely.

**4. Conclusions**

In conclusion, our vector-magnetometric microanalysis of the surface of a (001) magnetite single-crystal by spin-polarized low-energy electron microscopy shows that the surface magnetization is along two easy-axis [110] directions within the surface (001) plane. Our findings including substantial curvature of many of the domain walls and our interpretation of closure domain features including Néel-capped bulk Bloch walls and Bloch lines are in agreement with predictions based on micromagnetic simulations [41, 42].

**Acknowledgments**

This research was partly supported by the Spanish Ministry of Science and Innovation under Project No. MAT2009-14578-C03-01. Experiments were performed at the National Center for Electron Microscopy, Lawrence Berkeley National Laboratory, supported by the Office of Science, Office of Basic Energy Sciences, Scientific User Facilities Division, of the U.S. Department of Energy under Contract No. DE-AC02—05CH11231. A. T. N. acknowledges a Feodor Lynen Postdoctoral Fellowship from the Alexander von Humboldt Foundation.

# References

[1] A. A. Mills, The lodestone: History, physics, and formation, *Ann. Sci.* **61** (2004) 273.

[2] M. Friak, A. Schindlmayr, and M. Scheffler, Ab initio study of the half-metal to metal transition in strained magnetite, *New J. Phys.* **9** (2007) 5.

[3] M. I. Katsnelson, V. Yu. Irkhin, L. Chioncel, A. I. Lichtenstein, and R. A. de Groot, Half-metallic ferromagnets: From band structure to many-body effects, *Rev. Mod. Phys.* **80** (2008) 315.

[4] M. Bibes and A. Barthelemy, Oxide spintronics, *IEEE Trans. Elec. Dev.* **54**, (2007) 1003.

[5] E. Wada, K. Watanabe, Y. Shirahata, M. Itoh, M. Yamaguchi, and T. Taniyama, Efficient spin injection into GaAs quantum well across $Fe_3O_4$ spin filter, *Appl. Phys. Lett.* **96** (2010) 02510.

[6] M. Monti, B. Santos, A. Mascaraque, O. Rodríguez de la Fuente, M. A. Niño, T. O. Mentes, A. Locatelli, K. F. McCarty, J. F. Marco, and J. de la Figuera, Magnetism in nanometer-thick magnetite, *Phys. Rev. B* **85** (2012) 020404.

[7] R. M. Cornell and U. Schwertmann, The Iron Oxides, John Wiley & Sons Ltd, 1997.

[8] B. Stanka, W. Hebenstreit, U. Diebold, and S.A. Chambers, Surface reconstruction of $Fe_3O_4(001)$, *Surf. Sci.* **448** (2000) 49.

[9] R. Pentcheva, W. Moritz, J. Rundgren, S. Frank, D. Schrupp, and M. Scheffler, A combined DFT/LEED-approach for complex oxide surface structure determination: $Fe_3O_4(001)$, *Surf. Sci.* **602** (2008) 1299.

[10] S.A. Chambers, S. Thevuthasan, and S.A. Joyce, Surface structure of MBE-grown $Fe_3O_4(001)$ by x-ray photoelectron diffraction and scanning tunneling microscopy, *Surf. Sci.* **450** (2000) L273.

[11] A.V. Mijiritskii, M.H. Langelaar, and D.O. Boerma, Surface reconstruction of $Fe_3O_4(100)$, *J. Magn. Magn. Mat.* **211** (2000) 278.

[12] A.V. Mijiritskii and D.O. Boerma, The (001) surface and morphology of thin $Fe_3O_4$ layers grown by $O_2$-assisted molecular beam epitaxy. *Surf. Sci.* **486** (2001) 73.

[13] G. Mariotto, S. Murphy, and I. V. Shvets, Charge ordering on the surface of $Fe_3O_4(001)$, *Phys. Rev. B* **66** (2002) 245426.

[14] D. Stoltz, A. Önsten, U.O. Karlsson, and M. Göthelid, Scanning tunneling microscopy of Fe- and O-sublattices on (100), *Ultramicroscopy* **108** (2008) 540.

[15] G. S. Parkinson, N. Mulakaluri, Y. Losovyj, P. Jacobson, R. Pentcheva, and U. Diebold, Semiconductor half metal transition at the $Fe_3O_4(001)$ surface upon hydrogen adsorption, *Phys. Rev.* B **82** (2010) 125413.

[16] K. Jordan, A. Cazacu, G. Manai, S. F. Ceballos, S. Murphy, and I. V. Shvets, Scanning tunneling spectroscopy study of the electronic structure of $Fe_3O_4$ surfaces, *Phys. Rev.* B **74** (2006) 085416.

[17] R. Pentcheva, F. Wendler, H. L. Meyerheim, W. Moritz, N. Jedrecy, and M. Scheffler, Jahn-Teller stabilization of a "Polar" metal oxide surface: $Fe_3O_4(001)$, *Phys. Rev. Lett.* **94** (2005) 126101.

[18] G. S. Parkinson, Z. Novotny, P. Jacobson, M. Schmid, and U. Diebold, A metastable Fe(A) termination at the $Fe_3O_4(001)$ surface, *Surf. Sci.* **605** (2011) L42.

[19] M. Fonin, R. Pentcheva, Y. S. Dedkov, M. Sperlich, D. V. Vyalikh, M. Scheffler, U. Ruediger, and G. Guentherodt, Surface electronic structure of the $Fe_3O_4(100)$: Evidence of a half-metal to metal transition, *Phys. Rev.* B **72** (2005) 104436.

[20] J. G. Tobin, S. A. Morton, S. W. Yu, G. D. Waddill, I. K. Schüller, and S. A. Chambers, Spin resolved photoelectron spectroscopy of $Fe_3O_4$: the case against half-metallicity, *J. Phys.-Cond. Matt.* **19** (2007) 315218.

[21] M. Fonin, Y. S. Dedkov, R. Pentcheva, U. Ruediger, and G. Guentherodt, Magnetite: a search for the half-metallic state, *J. Phys.-Cond. Matt.* **19** (2007) 315217.

[22] M. Kurahashi, X. Sun, and Y. Yamauchi, Recovery of the half-metallicity of an $Fe_3O_4(100)$ surface by atomic hydrogen adsorption, *Phys. Rev.* B **81** (2010) 193402.

[23] Sangeeta Kale, S. M. Bhagat, S. E. Lo, T. Scabarozi, S. B. Ogale, A. Orozco, S. R. Shinde, T. Venkatesan, B. Hannoyer, B. Mercey, and W. Prellier, Film thickness and temperature dependence of the magnetic properties of pulsed-laser-deposited $Fe_3O_4$ films on different substrates, *Phys. Rev.* B **64** (2001) 205413.

[24] J Korecki, B. Handke, N. Spiridis, T. lzak, I. Flis-Kabulska, and J. Haber, Size effects in epitaxial films of magnetite, *Thin Solid Films* **412** (2002) 14.

[25] D. C. Kundaliya, S. B. Ogale, L. F. Fu, S. J. Welz, J. S. Higgins, G. Langham, S. Dhar, N. D. Browning, and T. Venkatesan. Interfacial characteristics of a $Fe_3O_4/Nb(0.5\%):SrTiO_3$ oxide junction, *J. App. Phys.* **99** (2006) 08K304.

[26] J. G. Zheng, G. E. Sterbinsky, J. Cheng, and B. W. Wessels, Epitaxial $Fe_3O_4$ on $SrTiO_3$ characterized by transmission electron microscopy, *J. Vac. Sci. Tech.* A **25** (2007) 1520.

[27] W. Eerenstein, L. Kalev, L. Niesen, T. T. M. Palstra, and T. Hibma, Magneto-resistance and superparamagnetism in magnetite films on MgO and $MgAl_2O_4$, *J. Mag. Mag. Mat.* **258** (2003) 73.

[28] A. Bollero, M. Ziese, R. Hohne, H.C. Semmelhack, U. Kohler, A. Setzer, and P. Esquinazi, Influence of thickness on microstructural and magnetic properties in $Fe_3O_4$ thin films produced by PLD, *J. Mag. Mag. Mat.* **285** (2005) 279.

[29] M. Fonin, C. Hartung, U. Rudiger, D. Backes, L. Heyderman, F. Nolting, A. Fraile Rodriguez, and M. Klaui, Formation of magnetic domains and domain walls in epitaxial $Fe_3O_4$(100) elements, *J. App. Phys.* **109** (2011) 07D315.

[30] Ö. Özdemir, S. Xu, and David J. Dunlop, Closure domains in magnetite, *J. Geophys. Res.* **100** (1995) 2193.

[31] A. Ambatiello, K. Fabian, and V. Hoffmann, Magnetic domain structure of multidomain magnetite as a function of temperature: observation by Kerr microscopy, *Phys. Earth Plan. Int.* **112** (1999) 55.

[32] Ö. Özdemir and D. J. Dunlop, Magnetic domain observations on magnetite crystals in biotite and hornblende grains, *J. Geophys. Res.* **111** (2006) 12.

[33] S. Foss, E. D. Dahlberg, R. Proksch, and B. M. Moskowitz, Measurement of the effects of the localized field of a magnetic force microscope tip on a 180 domain wall, *J. App. Phys.* **81** (1997) 5032.

[34] T.G. Pokhil and B.M. Moskowitz, Magnetic domains and domain walls in pseudo-single-domain magnetite studied with magnetic force microscopy, *J. Geophys. Res.* **102** (1997) 22681.

[35] R. J. Harrison, R. E. Dunin-Borkowski, and A. Putnis, Direct imaging of nanoscale magnetic interactions in minerals, *Proc. Nat. Acad. Sci.* **99** (2002) 16556.

[36] A. Hubert and R. Schafer, Magnetic domains: the analysis of magnetic microstructures, Springer, 1998.

[37] N. Rougemaille and A. K. Schmid, Magnetic imaging with spin-polarized low-energy electron microscopy, *Eur. Phys. J. App. Phys.*, **50** (2010) 20101.

[37] K. Grzelakowski, T. Duden, E. Bauer, H. Poppa, and S. Chiang, A new surface microscope for magnetic imaging, *IEEE Trans. Mag.* **30** (1994) 4500.

[38] T. Duden and E. Bauer. A compact electron-spin-polarization manipulator. *Rev. Sci. Inst.* **66** (1995) 2861.

[39] E. Bauer, Low energy electron microscopy, *Rep. Prog. Phys.* **57** (1994) 895.

[40] Mateck GmbH. www.mateck.de.

[41] W. Williams and T. M. Wright, High-resolution micromagnetic models of fine grains of magnetite, *J. Geophys.Res.* **103** (1998) 30537.

[42] S. Xu and D. J. Dunlop, Micromagneticmodeling of Bloch walls with Néel caps in magnetite, *Geophys. Res. Lett.* **23** (1996) 2819.

[43] C. E. Geiß, F. Heider, H. C. Soffel, Magnetic domain observations on magnetite and titanomaghemite grains (0.5–10 μm), *Geophys. J. Int.* **124** (1996) 75.

[44] T.G. Pokhil, B.M. Moskowitz, Magnetic domains and domain walls in pseudo-single-domain magnetite studied with magnetic force microscopy, *J. Geophys. Res.* **102** (1997) B10, 22.681

Figure 1. (a) Electron reflectivity versus energy for the magnetite (100) surface, integrated from a sequence of LEEM images acquired at different energies. (b) In-plane spin-dependent asymmetry in the reflectivity as a function of energy, obtained by integrating the reflected intensity from a uniform magnetic domain with opposite electron beam spin polarizations.

Figure 2. SPLEEM images of an area on the magnetite surface. The field of view is 12 μm, and the electron energy is 4.7 eV. The spin direction of the electron beam is out-of-the plane of the surface in panel (a) and in the plane of the image with azimuthal angles of 0° and 90° in panels (b) and (c), respectively. (d) Left: Color representation of the magnetization vector-field obtained from the combination of the previous SPLEEM images. In-plane angle, magnitude and z-component have been mapped on hue, saturation and lightness. Right shows the equidistant azimuthal projection of the color sphere. (e) Scatter plot of the observed in-plane magnetization vectors showing that the most common orientations of the magnetization are along two directions (biaxial easy axis). Each data point is colored as described in (d).

Figure 3. Collage of several SPLEEM images, surveying the magnetite surface. Each individual image has a field of view of 12 μm. (a) SPLEEM images with the electron beam spin direction in-plane along the surface [011] direction, corresponding roughly to the image diagonal (b) SPLEEM images in the same area with the electron beam spin direction along the orthogonal direction [01-1]. The electron energy is 4.7 eV.

Figure 4. Details of several particular magnetization configurations observed in the survey of domain structures (the regions shown in panels b, d, and e are from within the survey reproduced in Fig. 3). Small red arrows show the in-plane magnetization directions as measured from pairs of SPLEEM images acquired with reversed beam polarizations. (a) 90° wall (background image along the [110] direction, which shows a white domain). The image size is 2.7 μm. (b) Wavy Néel cap on a 180° wall (background image along [110] direction which shows no contrast at the domain wall). The image size is 5.3 μm. (c) Magnetization angle relative to the [110] (positive angles are counterclockwise, acquired along the yellow line indicated in previous panel): Within the domain wall the magnetization is along the [-110] suggesting that the structure is a 180° Néel-capped Bloch wall. (d) Bloch line separating two sections of opposite chirality within a Néel-capped 180° wall (background image along [1-10]). Bloch wall image size is 2.7 μm. (e) Periodic array of 90° walls, image size is 5.3 μm (the background image is along [100] so each domain appears white and black with the same intensity). (f) Profile of the magnetization angle relative to the [1-10] direction, along the yellow line marked in (e). Large red arrows indicate the average magnetization in each domain.

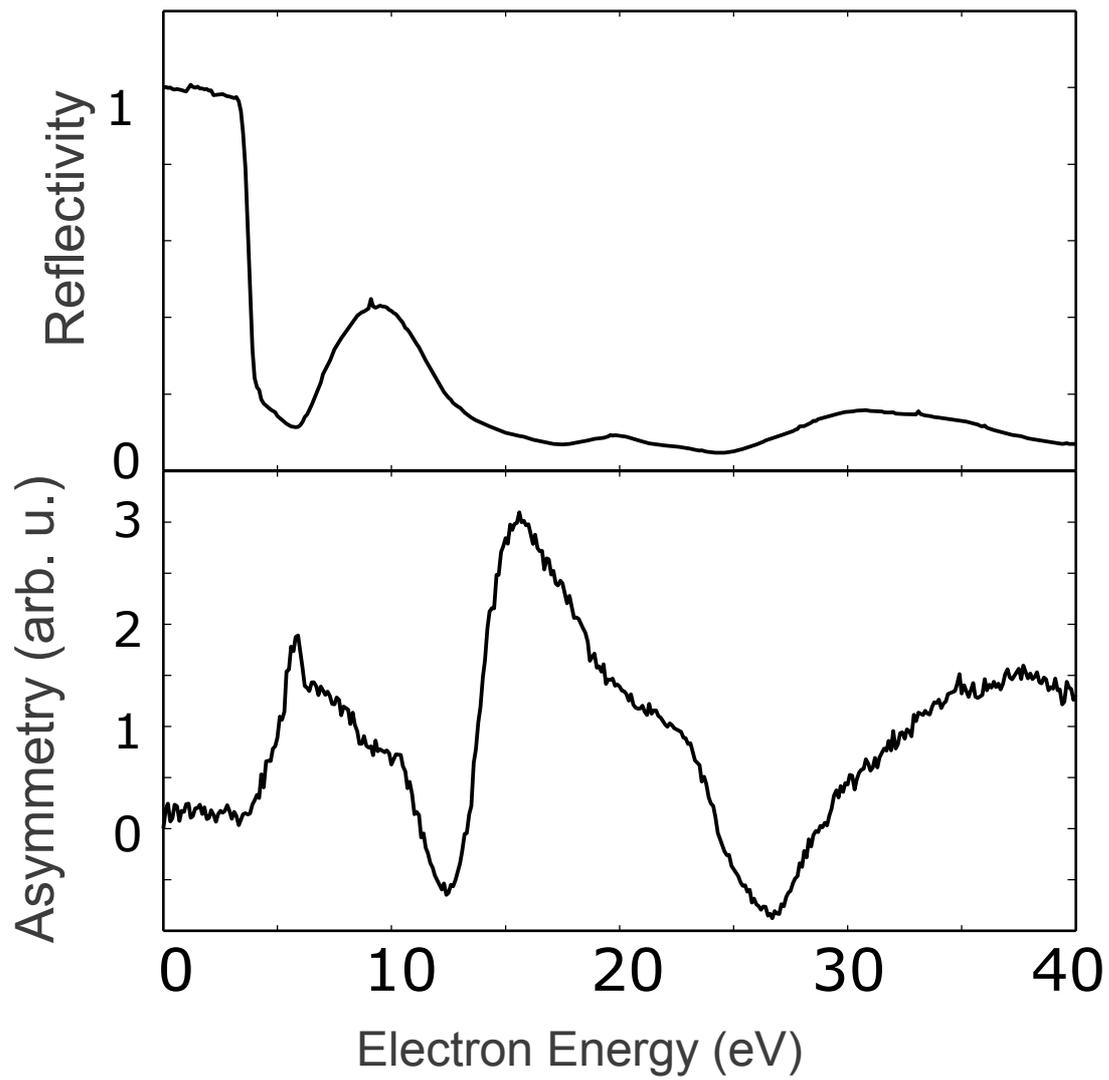

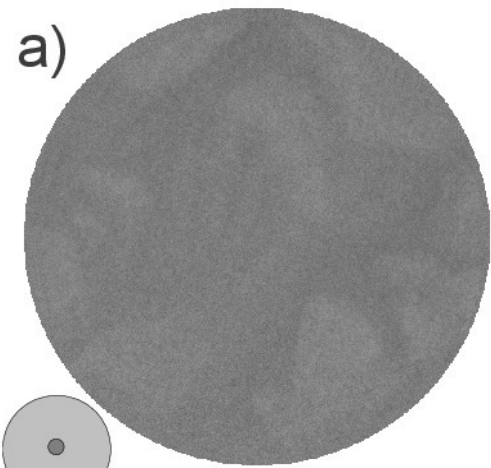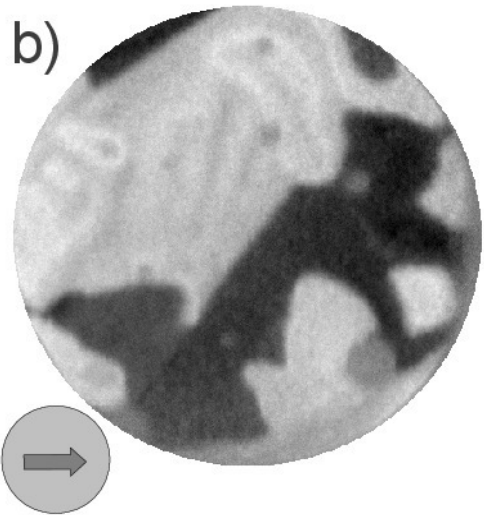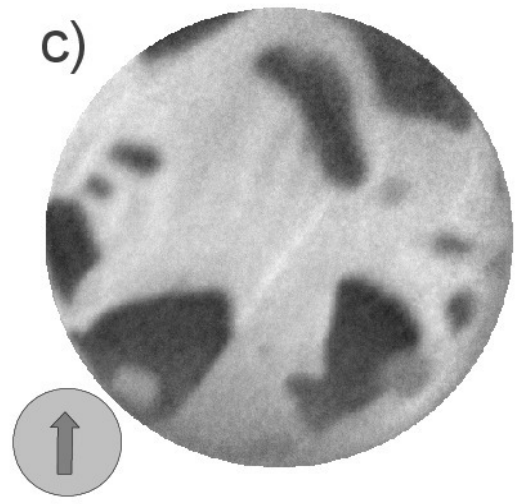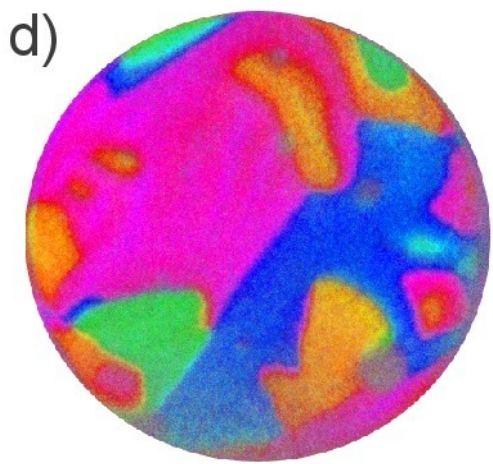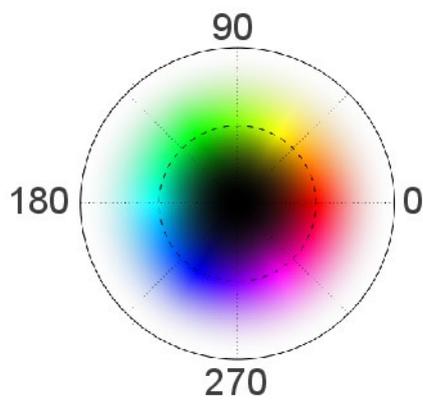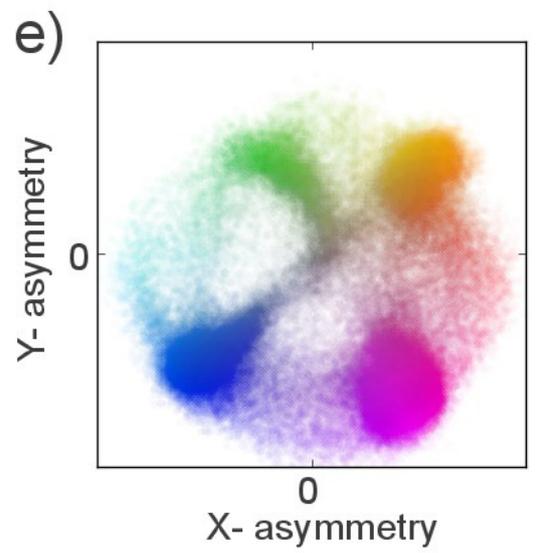

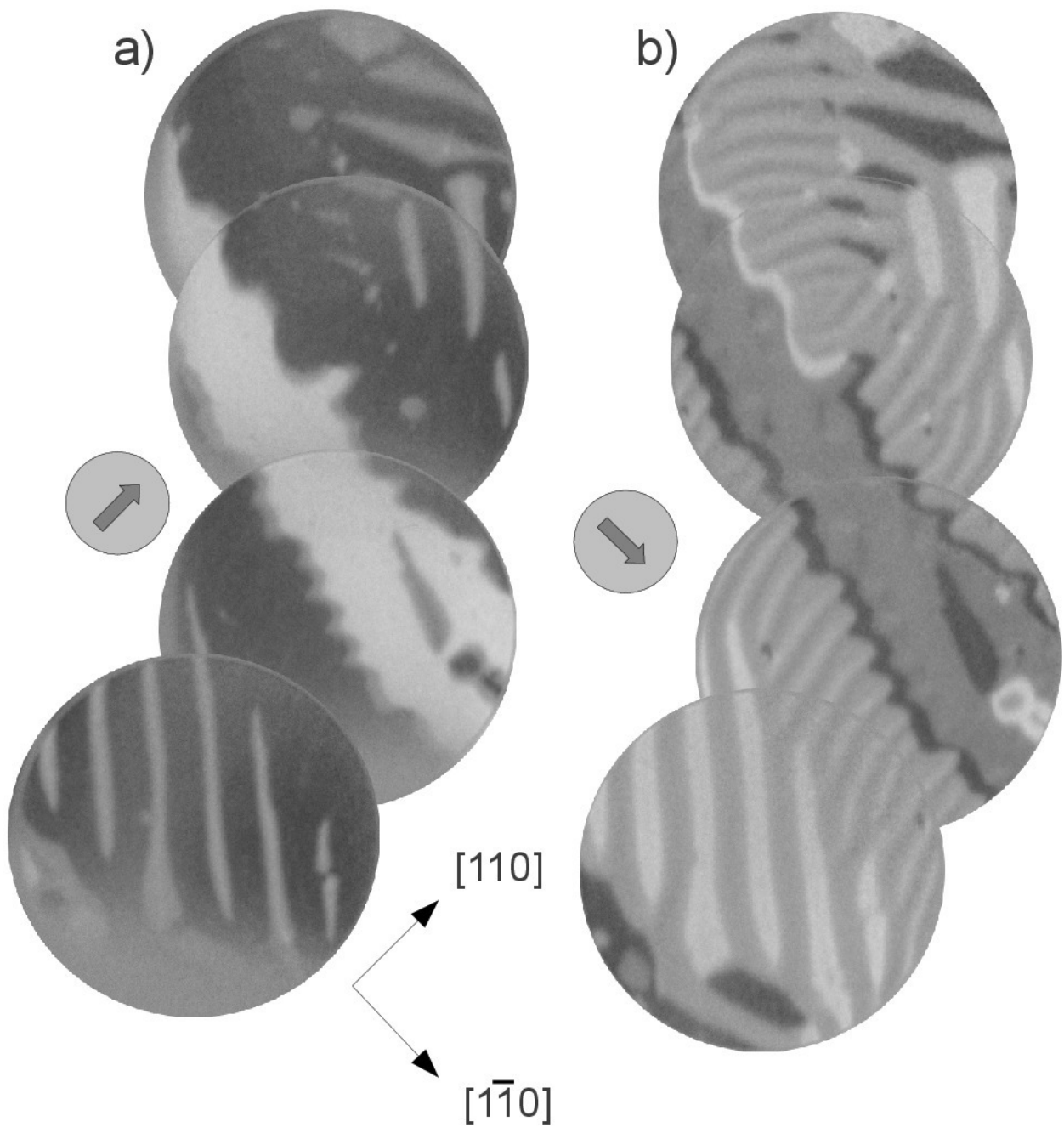

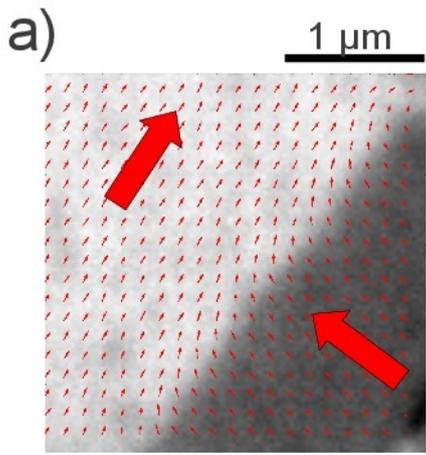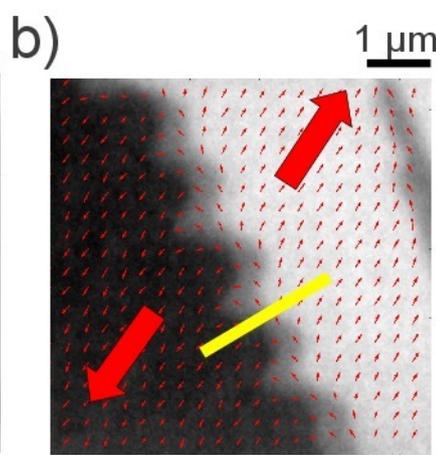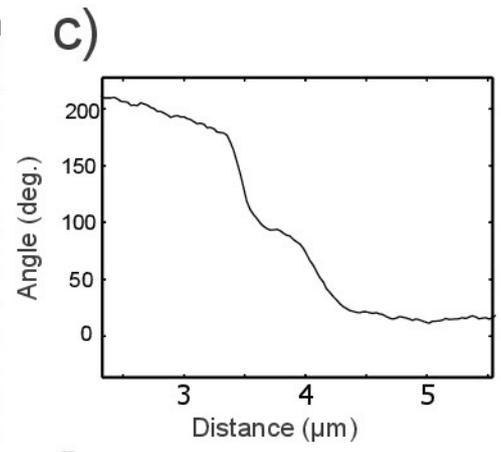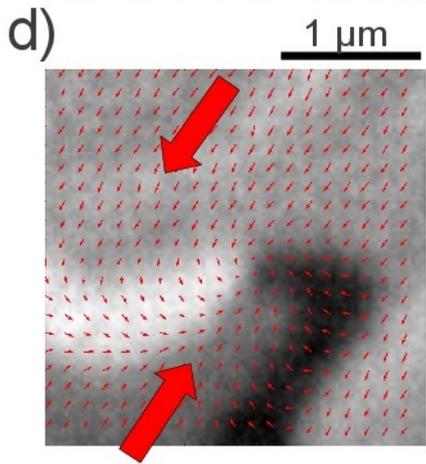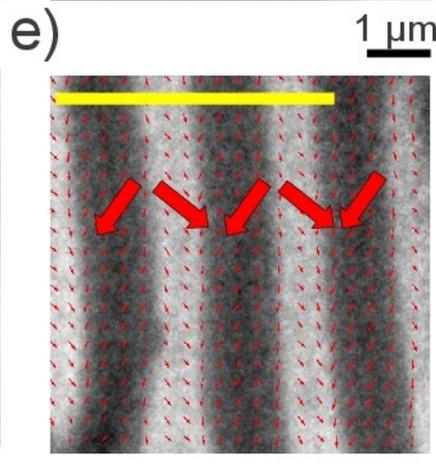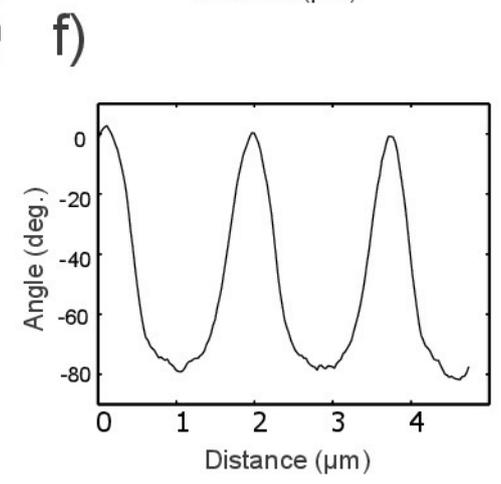